\documentclass[smallextended,runningheads]{svjour3}
\usepackage{amsmath,amssymb,amsfonts,amsbsy}

\usepackage[T1]{fontenc}
\usepackage{enumerate,url}

\usepackage{mathtools}
\usepackage[hidelinks]{hyperref}
\usepackage[nodoi]{apacite} 
\usepackage[british]{babel}
\usepackage{natbib}		
\setcitestyle{round,aysep={}}
\usepackage{xcolor}

\usepackage[utf8]{inputenc} 	
\usepackage[ruled]{algorithm2e}
\usepackage{comment}
\usepackage{tabularray}
\usepackage{tikz}
\usetikzlibrary{arrows.meta}
\usetikzlibrary{shapes.geometric}
\usepackage[normalem]{ulem}
 \usepackage{mathptmx}      % use Times fonts if available on your TeX system
\usepackage{latexsym}
\usepackage{graphicx}%
\usepackage{multirow}%
\usepackage{amsmath,amssymb,amsfonts}%
\usepackage{mathrsfs}%
\usepackage[title]{appendix}%
\usepackage{xcolor}%
\usepackage{textcomp}%
\usepackage{booktabs}%
\usepackage{comment}
\usepackage{lineno}
\newtheorem{ex}{Example}%

\newtheorem{assumption}{Assumption}

\newcommand{\vect}[1]{\left( #1_1, \dots, #1_N \right)^\top}
\newcommand{\wcor}{\rho^w} % w-correlation
\newcommand{\weighted}[2]{\langle #1,#2\rangle_w} % frobenius inner product
\newcommand{\weightednorm}[1]{\lVert #1\rVert_w} % frobenius norm
\newcommand{\Sidak}{{\v{S}}id{\'a}k }
\usepackage{lineno}
\newcommand{\summ}{U} % summand for the test statistic
\newcommand{\tstat}{T} % test statistic
\newcommand{\hp}{\mathcal{H}}

\journalname{}
\smartqed  % flush right qed marks, e.g. at end of proof
\usepackage{graphicx}
\begin{document}
\title{Utilizing Multiple Testing for Grouping in Singular Spectrum Analysis}
%\subtitle{Do you have a subtitle?\\ If so, write it here}

\titlerunning{Grouping in Singular Spectrum Analysis} 

\author{Maryam Movahedifar  \and Friederike Preusse \and Anna Vesely  \and Thorsten Dickhaus \footnote[2]{corresponding author}
}

\authorrunning{M. Movahedifar  \and F. Preusse \and A. Vesely \and T. Dickhaus} % if too long for running head

\institute{M. Movahedifar \at
	\email{movahedm@uni-bremen.de}\\
	Institute for Statistics, University of Bremen, 28344 Bremen, Germany \and
	F. Preusse \at
	\email{preusse@uni-bremen.de}\\
	Institute for Statistics, University of Bremen, 28344 Bremen, Germany \and
	A. Vesely \at
	\email{anna.vesely2@unibo.it}\\
	Department of Statistical Sciences, University of Bologna, 40126 Bologna, Italy\and
	T. Dickhaus \at
	\email{dickhaus@uni-bremen.de}\\
	Institute for Statistics, University of Bremen, 28344 Bremen, Germany 
}

\date{Received: date / Revised: date}
%
% The correct dates will be entered by the editor
\maketitle
\begin{abstract} %is supposed to be at most 150 words, currently: 119 words
A key step in separating signal from noise in time series by means of singular spectrum analysis (SSA) is grouping. We present a multiple testing method for the grouping step in SSA. As separability criterion, we utilize the weighted correlation between the signal and the noise component of the (reconstructed) time series, and we test whether this weighted correlation is equal to zero. This test has to be performed for several possible groupings, resulting in a multiple test problem. The null distributions of the corresponding test statistics are approximated by a wild bootstrap procedure. The performance of our proposed method is assessed in a simulation study, and we illustrate its practical application with an analysis of real world data. 

\end{abstract}
\newpage

\section{Introduction}\label{sec1}

Pre-processing data is an essential step for data analysis. In this regard, noise reduction and signal extraction are considered an important step of data analysis in all fields of study; see, among many others, \cite{Pyle1999,OLIVERI20199,Maryam2}. In the present work, we are concerned with the statistical analysis of noisy time series. 
Singular Spectrum Analysis (SSA) is a non-parametric denoising method for analyzing time series. SSA is used in a wide range of application areas, including ecology and environmental research, medicine, economics, and finance (see, for example, \cite{ex1, ex2, ex3, ex4, ex5, ex6, ex7, ex8, ex9, Bio-2015} and references therein). A detailed description of other popular denoising methods for one-dimensional time series is provided by \cite{kohler2005comparison}. 

In the SSA method, the observed time series (modeled as signal plus noise) is transformed into a matrix $\mathbf{X}$, the columns of which consist of overlapping consecutive data points. This matrix $\mathbf{X}$ is called the trajectory matrix. The key step in SSA, when used for denoising, is grouping. The grouping step results in a decomposition of $\mathbf{X}$. Choosing the number of groups equal to two allows for applying a grouping criterion which is targeted towards separating signal components from noise components. We will provide more details in Section \ref{section.intssa}. General introductions to SSA have been provided by \cite{Golyandina2001}, \cite{golyandina2018singular}, and \cite{Golyandina}.

Traditional approaches for grouping in SSA are often subjective. These approaches typically involve selecting a group size based on expert knowledge or heuristic rules. A brief overview of some automatic grouping approaches is provided in Section \ref{sec-related-work}. In the present work, we propose a new method for automatic grouping, which relies on multiple statistical hypotheses testing. Thus, the proposed method provides statistical guarantees for the resulting grouping, which is in contrast to all existing approaches that we are aware of. 
As separability criterion, we utilize the weighted correlation ($w$-correlation for short) between the signal and the noise component of the (reconstructed) time series. For optimal denoising, this $w$-correlation should be equal to zero, and we develop multiple tests for testing the null hypothesis of zero $w$-correlation for several possible groupings simultaneously; see Section \ref{seq:wbdd_proc} for details. A wild bootstrap procedure proposed by \cite{hounyo2023wild} is employed to approximate the null distribution of the empirical $w$-correlation. 

%%%%%%%%%%%%%%%%%%%%%%%%%%%%%%%%%%%%%%%%%%%%%%%%%%%%%
%              S E C T I O N     1.1                %
%%%%%%%%%%%%%%%%%%%%%%%%%%%%%%%%%%%%%%%%%%%%%%%%%%%%%
\subsection{Related work}\label{sec-related-work}
\paragraph{Automatic grouping}
Generally, the sub-signals (or components) of a time series which correspond to the signal are the $g$ leading components of the time series. In this, the term ``leading'' corresponds to the order of the eigenvalues of $\mathbf{X} \mathbf{X}^\top$, where $^\top$ denotes matrix transposition throughout; see the description around Equation (2.2) in \cite{Golyandina} and the explanations at the end of Section 2.1.2.3 in \cite{Golyandina}. Therefore, determining the grouping in SSA corresponds to finding the group size $g$. If no prior information about $g$ is available, methods for automatic grouping can be applied. 

If there is no information about the rank of the signal, the $w$-correlation can be utilized. For example, \citet{Bilancia2010} apply hierarchical clustering to the $w$-correlation matrix, while \citet{Alonso2008} utilize a $k$-means clustering procedure.
If specific time series characteristics are of interest, one can evaluate different values of $g$ by comparing the estimated characteristics of the corresponding reconstructed signals with the observed characteristics. For example, if the aim of SSA is forecasting, one can apply SSA to historical data and compare the accuracy of the forecasts based on different values of $g$ with already observed data. Then, $g$ is chosen such that the corresponding forecast has the best accuracy; see Page 66 of \cite{Golyandina}.

If the signal is assumed to have finite rank, subspace-based approaches can be used.
In this case, determining $g$ corresponds to estimating the rank of the signal. This can be done using information criteria, which was first proposed by \citet{Wax1985} \citep[see ][for an overview]{Stoica2004}. If the signal is assumed to be a sum of complex exponentials, automatic estimation of the signal rank can be done based on the rotational invariance principle \citep{Roy1989, Badeau2004, Papy2007} or the shift-invariance principle \citep{Albert2023}. Estimation of the model rank using deep learning models has been proposed by \citet{Moon2021} for signals following an autoregressive moving-average (ARMA) process.
An overview of the relation between subspace-based methods and SSA is given in Chapter 3.8 of \citet{Golyandina2020}.

Procedures based on $w$-correlation and subspace-based methods become unreliable if the signal and noise are not well separated \citep{Golyandina2020}. Furthermore, they offer no statistical guarantees that the determined group of signal components contain only true signal components of the time series. In some settings, it is important to be confident that the estimated signal does not contain any noise. Hence, in such settings statistical guarantees are of interest. The procedure to estimate $g$ proposed in this work offers such guarantees. The proposed method is based on the $w$-correlation and is therefore applicable to a wide range of time series, which is in contrast to the subspace-based methods.

Note that variations of SSA exists which do not require to determine a grouping, see for example \citet{Yang2025, Bogalo2021}. However, in this work, we are interested in the classical framework of SSA which requires grouping.

\paragraph{Bootstrap procedures for time series}
When applying resampling procedures to time series, the temporal dependence of the series need to be accounted for. One family of resampling procedures for dependent data are block-based bootstrap procedures, which resample blocks instead of single observations. Both \citet{Kunsch1989} and \citet{Liu1992} introduced the moving block bootstrap, a procedure which has been extended and adapted to different settings, see for example \citet{carlstein1986use, politis1994stationary, paparoditis2001tapered, Donald2004,Paparoditis2011}. An alternative to block-based bootstrap procedures are wild bootstrap procedures for dependent data \citep[e.g.,][]{shao2010dependent, Leucht2013, hounyo2023wild}. In contrast to block-based methods, wild bootstrap methods for dependent data do not require the partitioning of the data into blocks. Instead, they utilize auxiliary variables that are dependent in a way that captures the (auto-)dependence in the original time series. Like the traditional wild bootstrap, introduced by \citet{Wu1986} and \citet{Hardle1993}, wild bootstrap procedures for dependent data are applicable to heterogeneously distributed data. 

For overviews and comparisons of different bootstrap procedures for dependent data, we refer to \citet{Lahiri2003} and \citet{Kreiss2011}.

\subsection{Main contributions}

	The weighted correlation coefficient, or $w$-correlation, is a measure that can be used for judging how well the signal and noise have been separated in SSA. A value close to zero is an indication that the number of groups used is sufficient and hence that noise and signal are well separated. To determine the number $g$ in the grouping step of SSA, we conduct inference on the weighted correlation coefficient utilizing the wild bootstrap procedure proposed by \cite{hounyo2023wild}. This procedure allows us to approximate the (analytically intractable) sampling distribution of the empirical $w$-correlation. We utilize this approximation to make inference on the $w$-correlation. To be specific, we test for all possible groupings if the resulting $w$-correlation is significantly different from zero. To account for the multiplicity of the testing problem, we control the family-wise error rate (FWER). Therefore, the probability of making at least one false rejection is bounded from above by a pre-defined significance level, which we will denote by $\alpha$. In other words, we control the probability that noise is included in the reconstructed signal. This is a novelty, since established automatic grouping procedures cannot give such guarantees. Since the proposed method is based on the $w$-correlation, it is applicable to time series with finite or infinite rank.

\subsection{Overview of the rest of the material}

	The rest of this material is organized as follows. A description of the SSA methodology,  including notations and assumptions, is given in 
	Section \ref{section.ssa}. A description of the proposed procedure is given in Section \ref{seq:wbdd_proc}. Simulation studies and real world data analysis that illustrate the application of the proposed methodology are given in Sections \ref{simulation} and \ref{real_world_examples}, respectively. Finally, we conclude in Section \ref{conclusion} with a discussion on our findings and directions for future research.

%%%%%%%%%%%%%%%%%%%%%%%%%%%%%%%%%%%%%%%%%%%%%%%%%%%%%%%%%%%%
%        S E C T I O N     2                               %
%%%%%%%%%%%%%%%%%%%%%%%%%%%%%%%%%%%%%%%%%%%%%%%%%%%%%%%%%%%%
\section{Singular Spectrum Analysis}\label{section.ssa}
The origins of SSA can be traced back to the work of \citet{BROOMHEAD1986217}. SSA, as a branch of time series analysis, is a nonparametric approach that can be used for denoising a time series. Furthermore, the SSA method is capable of filtering and forecasting time series. It can be carried out as a univariate or as a multivariate method. 
The major goal of the SSA method is  to decompose the initial ordered series (e.\ g., a time series) into a sum of  different components which can be considered as either a trend, a periodic, a quasi-periodic (perhaps amplitude-modulated), or a noise element. We refer to 
\cite{MOVAHEDIFAR201852,  SILVA2019134, article35, HASSANI38, article39} for more details. In the following subsections, descriptions of SSA and of separability concepts are presented. For more information, see \cite{Bio-2015}.

\subsection{A Brief Description of Basic SSA}\label{section.intssa}

Let $\textbf{Y}_N=\vect{y}$ denote a (univariate) time series of length $N$. Fix an integer $\textit{L}$, which is called window length, where ~$2\leq \textit{L} \leq N/2$. Then, basic SSA consists of the following four steps:
\begin{enumerate}
	\item \textbf{ Embedding}: \label{embed}
	This step transforms the $\textbf{Y}_N$ into the multi-dimensional series $\mathbf{X}^{(1)},$ $\ldots, \mathbf{X}^{(K)}$ with vectors $\textbf{X}^{(i)} =\left(y_i,\ldots, y_{i+L-1}\right)^{\top} \in\mathbb{R}^{L}$, where $K = N - L +1$. The vectors $\mathbf{X}^{(i)}$ are called $L$-lagged vectors (or, simply, lagged vectors). The embedding step includes only one parameter, namely, the window length 
	$L$. The purpose of this step is to form the trajectory matrix $\textbf{X} =\left[\mathbf{X}^{(1)},\ldots, \mathbf{X}^{(K)}\right] \in \mathbb{R}^{L \times K}$. Depending on several criteria like complexity of the data, the purpose of the analysis of data and, in the context of prediction, the forecasting horizon, the parameter $L$ can be chosen. \cite{Bio-2015} recommend that $L$ should be ‘large’, but not larger than $N/2$.  In order to find a suitable value of $L$, the concept of separability can be considered; cf. Section \ref{separability}. By definition, an $(L \times K)$-matrix $\mathbf{Z}$ is a Hankel matrix if its elements $z_{ij}$ are constant on the ``matrix diagonals'', i.\ e., $z_{ij} \equiv \mathfrak{z}_k$ for all $(i, j)$ such that $i + j = k$, for any $2 \leq k \leq L + K$. It is worthwhile to notice that any trajectory matrix is a Hankel matrix with $\mathfrak{z}_k = y_{k-1}$, $2 \leq k \leq L + K$.
	\item \textbf{Singular Value Decomposition (SVD)}:
	In this step, the trajectory matrix $\mathbf{X}$ is decomposed into a sum of
	rank-one elementary matrices. Let $\lambda_1,\ldots,\lambda_L$ denote the eigenvalues of
	$\mathbf{X}{\mathbf{X}}^{T}$, which are ordered 
	in decreasing magnitude such that
	$\lambda_1\geq\dots\geq\lambda_L\geq 0$, and let
	$\mathbf{U}_1, \ldots, \mathbf{U}_L$ denote the eigenvectors of 
	$\mathbf{X}{\mathbf{X}}^{\top}$ corresponding to these eigenvalues.  Letting
	$d = \textrm {max} \{i: \lambda_i> 0 \} = \text{rank}(\textbf {X}) $, 
	the SVD of the trajectory matrix can be written as $\mathbf{X}=\mathbf{X}_1+\cdots+\mathbf{X}_d $, where  $\textbf {X}_i ={\sqrt\lambda_i} \mathbf{U}_i {\mathbf{V}_i}^{\top}$ and $\mathbf{V}_i=\textbf {X}^{\top} \mathbf{U}_i/{\sqrt\lambda_i}$ for $i=1,\ldots,d$. This SVD is optimal in the sense that among all the matrices $\mathbf{X}^{(g)}$ of rank $g<d$ the matrix $\sum_{i=1}^{g}\mathbf{X}_i$ minimizes the Frobenius norm 
    $\lVert \mathbf{X} -  \mathbf{X}^{(g)} \rVert_F$ of 
    $\mathbf{X} -  \mathbf{X}^{(g)}$, which is also sometimes referred to as the $L_2$-norm. Note that ${\lVert \mathbf{X} \rVert}_F^{2}=\sum_{i=1}^{d}\lambda_i$ and ${\lVert \mathbf{X}_i \rVert}_F^{2}=\lambda_i$ for $i=1,\ldots,d$. Thus, we can consider the ratio ${\lambda_i}/{\sum_{i=1}^{d}\lambda_i}$ as a quantitative measure of the contribution of the matrix $\mathbf{X}_i$ to the expansion $\mathbf{X}=\mathbf{X}_1+\cdots+\mathbf{X}_d $. Accordingly, ${\sum_{i=1}^{g}\lambda_i}/{\sum_{i=1}^{d}\lambda_i}$, the ratio of the sum of the first $g$ eigenvalues over the sum of all $d$ non-zero eigenvalues, is a quantitative measure of the optimal approximation of the trajectory matrix by matrices of rank $g$. Selecting an appropriate value of $g$ is very important. On the one hand, by selecting $g$ smaller than the true number of non-zero eigenvalues, some parts of the signal(s) will be lost. On the other hand, if one takes $g$ greater than the value that it should be, then noise components will be included in the reconstructed series, which one may want to avoid in certain applications.
	
	\item \textbf{Grouping}\label{grouping}:
	In this step, the set of indices $\{1,\ldots,d\}$ is partitioned into $\gamma$ disjoint subsets ${I}_1,\ldots, {I}_\gamma$. Let ${I}=\{i_1,\ldots,i_p\} \subset \{1, \ldots, d\}$ be such a subset (of size $p$). Then, the matrix $\mathbf{X}_I$ corresponding to the group $I$ is given by $\mathbf{X}_I=\mathbf{X}_{i_1}+\cdots+\mathbf{X}_{i_p}$.
	For example, if ${I}=\{2,3,5\}$, then $\mathbf{X}_I=\mathbf{X}_{2}+\mathbf{X}_{3}+\mathbf{X}_{5}$. Considering the SVD of $\mathbf{X}$, the split of the set of indices $\{1,\ldots,d\}$ into the disjoint subsets
	$I_1,\ldots,I_\gamma$ can be represented as 
    \begin{equation}\label{splitted-trajectory} 
    \mathbf{X}=\mathbf{X}_{I_1}+\cdots+\mathbf{X}_{I_\gamma}.
    \end{equation}
    If the original series is modeled as signal plus noise, one considers $\gamma = 2$ groups of indices, namely $I_1 = \{1,..., g\}$ and $I_2 = \{g + 1,..., L\}$.  The group $I_1$ is associated with the signal component and the group $I_2$ with noise. At the grouping step, diagnostic tools for differentiating between noise and signal are the periodogram, pairwise scatterplots of eigenvectors, or the eigenvalue spectrum, among others; cf. Section 3.3 of \cite{hassani2007}. Once we have selected the eigenvalues corresponding to signal and noise, respectively, we can evaluate the effectiveness of this selection via the notion of separability; see Section \ref{separability}. 
	
	\item \textbf{Diagonal Averaging}:
	The aim of this step is to transform every obtained matrix $\mathbf{X}_{I_j}$ from the grouping step into a Hankel matrix so that these can subsequently be converted back into a (reconstructed) time series.  In basic SSA, Hankelization of $\mathbf{X}_{I_j}$ is achieved via diagonal averaging of its elements over all $(i, j)$ such that $i+j = \text{const.}$ By performing the diagonal averaging of all matrix components $\mathbf{X}_{I_j}$ in the expansion of $\mathbf{X}$ from \eqref{splitted-trajectory}, we obtain another expansion: $\mathbf{X} = \tilde{\mathbf{X}}_{I_1} + . . . + \tilde{\mathbf{X}}_{I_\gamma}$, where $\tilde{\mathbf{X}}_{I_j}$ is the Hankelized version of the matrix $\mathbf{X}_{I_j}$. This is equivalent to the decomposition of the initial series $\textbf{Y}_N=\vect{y}$ into a sum of $\gamma$ reconstructed series, namely $y_t = \sum_{j=1}^{\gamma}\tilde{\mathbf{y}}_t(j)$ for each $t \in \{1, \ldots, N\}$, where $\tilde{\mathbf{y}}_t(j)$ corresponds to the matrix $\tilde{\mathbf{X}}_{I_j}$ for each $t \in \{1, \ldots, N\}$. In this, the operation of back-transforming a Hankel matrix into an ordered univariate series is referred to as reconstruction.
\end{enumerate}

\subsection{Separability}\label{separability}

In the context of SSA, the notion of separability is employed to assess the quality of the decomposition of $\textbf{Y}_N$ into signal components and noise components. Here, we focus on the weighted scalar product and the weighted correlation, respectively, as separability criteria. Consider a given decomposition $\textbf{Y}_N = \textbf{S} + \textbf{Z}$, where $\textbf{S} =(S_{1},\ldots,S_{N})^\top$ corresponds to the signal components and $\textbf{Z} =(Z_{1},\ldots,Z_{N})^\top$ represents the noise component. For each $t \in \{1, \ldots, N\}$, denote by $w_t = \min\{t, L, N-t+1\}$ the number of times that the element $y_t$ appears in the trajectory matrix of $\textbf{Y}_N$. Then, the weighted scalar product of $\mathbf{S}$ and $\mathbf{Z}$ is defined as
\[
\langle\mathbf{S}, \mathbf{Z}\rangle_{{w}} = \sum_{t=1}^{N}w_tS_tZ_t.
\]
Correspondingly, the (empirical) weighted correlation (${w}$-correlation) of $\mathbf{S}$ and $\mathbf{Z}$ is given by
\begin{equation}\label{def:wcorr}
	\wcor(\textbf{S},\textbf{Z}) = \frac{\weighted{\textbf{S}}{\textbf{Z}}}{\weightednorm{\textbf{S}} \cdot \weightednorm{\textbf{Z}}},
\end{equation}
where $\weightednorm{\cdot}$ denotes the norm induced by $\langle\cdot, \cdot\rangle_{{w}}$.

The absolute value of the ${w}$-correlation indicates the level of separability. Namely, a ${w}$-correlation close to zero means that the corresponding series are nearly ${w}$-orthogonal, and this is what one would expect in the case of a successful decomposition of $\textbf{Y}_N$ into signal components and noise components. In contrast, a large ${w}$-correlation (in absolute value) suggests that the two series are far from being ${w}$-orthogonal. Visualizing the matrix of absolute $w$-correlations between series components in a grayscale format is a graphical tool to find a plausible value of $g$. In such a graph, low $w$-correlations appear as white, while $w$-correlations close to one are depicted in black. This visual approach provides information about the relationships within the series components. For further details, we refer to Section 2.2.3.4 in \cite{Bio-2015}.

Our objective in the next section is to choose a grouping with (minimal) grouping index $g\in\{1,\ldots,d\}$ that leads to a separation of signal and noise in terms of a statistical criterion regarding the $w$-correlation. Namely, we propose to test for each candidate value of $g$ the resulting $w$-correlation of reconstructed signal and reconstructed noise for being equal to zero. The null distribution of the tests will be approximated based on the wild bootstrap approach for dependent data (WBDD) by \cite{hounyo2023wild}. 

%%%%%%%%%%%%%%%%%%%%%%%%%%%%%%%%%%%%%%%%%%%%%%%%%%%%%%%%%%%%%%%%%%
%                          S E C T I O N    3                    %
%%%%%%%%%%%%%%%%%%%%%%%%%%%%%%%%%%%%%%%%%%%%%%%%%%%%%%%%%%%%%%%%%%
\section{Identification of the grouping index}
\label{seq:wbdd_proc}

Suppose that we perform SSA on a time series $\textbf{Y}_N=\vect{y}$  with a certain window length $L\leq N/2$, obtaining $d\leq L$ non-null singular values. For any grouping index $g\in\{1,\ldots,d\}$ 
we say that the main signal and noise are separable if the $w$-correlation given in Eq.~\eqref{def:wcorr} is close to zero, i.e., if the numerator
\begin{align}
\sum_{t=1}^N \summ_{gt},\qquad \summ_{gt}=w_t S_{gt} Z_{gt} \label{eq:ti}
\end{align}
is close to zero, where (with slight abuse of notation) $\mathbf{S}_g$ and $\mathbf{Z}_g$ denote the reconstructed signal and noise components, respectively, for grouping index $g$. Our main goal is finding the best grouping, i.e., the smallest index $\hat{g}$ such that the main signal and noise are separable for all $g\in\{\hat{g},\ldots,d\}$. Notice that we always have separability when $g=d$, as we obtain $\mathbf{Y}=\mathbf{S}_d$ and $\mathbf{Z}_d=\mathbf{0}$, and so
\begin{align*}
\summ_{dt}=0\text{ for all }t\in\{1,\ldots,N\}\qquad\Longrightarrow\qquad \sum_{t=1}^N \summ_{dt}=0. %\label{eq:d_value}
\end{align*}
For any other grouping $g\in\{1,\ldots,d-1\}$, we are interested in testing
\begin{align}
\hp_g:\, \mathbb{E}\left[\sum_{t=1}^N \summ_{gt}\right]=0,
\label{eq:nullhp0}
\end{align}
against a two-sided alternative, where $\mathbb{E}$ refers to the true, but unknown data-generating process. If $\hp_g$ is true, then the grouping with index $g$ leads to separability. Otherwise, the resulting main signal and noise are not separable, and we should use a larger grouping index. Therefore, we simultaneously test the null hypotheses\\ 
$\left(\hp_g: g\in\{1,\ldots,d-1\}\right)$ with control of the FWER. Then, we define the selected grouping index as
\begin{align}
\hat{g}=\max\{g\in\{1\ldots,d-1\}\;:\;\hp_g\text{ is rejected}\}+1 \label{def:gstar}
\end{align}
if the maximum exists, and $\hat{g}=1$ otherwise.

For given $g$, we conduct a bootstrap hypothesis test for $\hp_g$ based on wild bootstrapping, accounting for the dependence among $\summ_{g1},\ldots,\summ_{gN}$. To this end, we first give a short overview of general bootstrap hypothesis tests. We then present the procedure of generating wild bootstrap samples for testing $\hp_g$, as well as the computation of the test statistic and corresponding p-value.
Following this, we intend to examine all null hypotheses $\hp_1,\ldots,\hp_d$ simultaneously to determine the grouping index $\hat{g}$. \\

\subsection{Bootstrap Hypothesis Tests}\label{seq:general_bootstrap_testing}
To test a hypothesis using bootstrapping, $B$ bootstrap samples, for which the null hypothesis is true and which consist of $\mathcal{N}$ pseudo-observations each, are generated. For each bootstrap sample $b=1,\ldots, B$ the test statistic $T_b$ is computed \citep{MacKinnon2009}. 
Then, the bootstrap p-value for a two-sided null hypothesis is defined as 
\begin{equation}
p_{boot}=\frac{\sum_{b=1}^{B}\mathbf{1}(|T_{b}| \geq |T_{obs}|)+1}{B+1},
\label{eq:p_boot}
\end{equation}
where $T_{obs}$ corresponds to the test statistic based on the observed sample and $\mathbf{1}(\cdot)$ denotes the indicator function. Thus, the bootstrap p-value $p_{boot}$ represents the proportion of test statistics that are at least as extreme than $T_{obs}$ amongst all test statistics \citep{MacKinnon2009}. This accentuates the requirement that the bootstrap samples are obtained such that they follow the null hypothesis. It also indicates that we do not need to make any assumptions about the distribution of the test statistic $T_{obs}$. However, it is important that the bootstrap data generating process is valid, i.e., in our case accounts for the dependency and non-stationarity in the data. Thus, we use the wild bootstrapping procedure for dependent data (WBDD) introduced by \cite{hounyo2023wild}.

\subsection{Wild Bootstrap for SSA}\label{seq:WBDD_for_SSA}
The framework of the WBDD permits general dependence conditions and data heterogeneity. WBDD combines the ideas of wild bootstraps and tapered block bootstrap.

In the context of SSA, we utilize WBDD to estimate the sampling distribution of $\mathbf{\summ}_g=(\summ_{g1},\ldots,\summ_{gN})^\top$ as in \eqref{eq:ti}. To this end, we consider the centered time series $\mathbf{R}$ with $R_t\coloneqq U_{gt}-\Bar{\mathbf{\summ}}_g$ with $\Bar{\mathbf{\summ}}_g=N^{-1}\sum_{t=1}^N U_{gt}$.  A sequence of data-tapering windows $v_n(t)\in[0,1]$, $v_n(t)=0$ for $t\notin\{1,\ldots,n\}$ is used as weights. Denote by $||v_n||_1$ and $||v_n||_2$ the $L_1$ and $L_2$-norm, respectively.

To generate a WBDD sample, we first compute the tapered moving block sample mean \begin{equation*}
	\Bar{R}_{\ell, v}=\frac{1}{Q}\sum_{j=1}^Q\sum_{i=1}^\ell\frac{v_\ell(i)}{||v_\ell||_1}R_{i+j-1},
\end{equation*} with $\ell\in\{1,\ldots,N\}$ denoting the (fixed) block size and $Q=N-\ell+1$. A WBDD pseudo observation $R^\star$ is is computed by
\begin{equation*}
R^\star_t=(R_t-\Bar{R}_{\ell,w})\eta_t+\Bar{R},
\end{equation*}
with \begin{equation*}
	\eta_t\coloneq \sum_{j=1}^Q\frac{v_\ell(t-j+1)}{||v_\ell||_2}\sqrt{\ell}a_j,
\end{equation*} where $(a_j)_{j\in\{1,\ldots,Q\}}$ denotes a sequence of i.i.d. random variables.

As in \citet{hounyo2023wild}, we consider \begin{equation}
\label{eq:w}
v_n(t)=v((t-0.5)/n), 
\end{equation} where $v(\cdot)$ is a single function $v:\mathbb{R}\to[0,1]$. The function $v(\cdot)$ should meet three assumptions \citep{hounyo2023wild}. To maintain a self-contained presentation, we repeat the assumptions here.
\begin{assumption}
We have $v(t)\in[0,1]$ for all $t\in\mathbb{R}$, $v(t)=0$ if $t\notin[0,1]$, and $v(t)>0$ for $t$ in a neighborhood of $1/2$.
\end{assumption}
\begin{assumption}
    The function $v(t)$ is symmetric around $t=1/2$ and nondecreasing for $t\in[0,1/2]$.
\end{assumption}
\begin{assumption}

    The self-convolution is twice continuously differentiable at the point $t=0$, where $v\ast v(t)=\int_{-1}^1v(x)v(x+|t|)dx$.
\end{assumption}

Additional assumptions about the sequence $(a_j)_{j\in\{1,\ldots,Q\}}$ are made \citep{hounyo2023wild}. Once again, we repeat the assumptions here.
\begin{assumption}
   The sequence of random variables $(a_j)$, $j= 1,\ldots,Q=N-\ell+1$ is independent of the original observed sample $\mathbf{\summ}_g$. Moreover, it is independent and identically distributed and satisfies the following regularity conditions: $\mathbb{E}(a)=0$, $Var(\sqrt{\ell}a)=\ell\mathbb{E}(a)^2\to 1$ and for some $\delta>0$, $\ell\mathbb{E}(|a|^{2+\delta})\to C_\delta<\infty$ as $N\to\infty$, $\ell\to\infty$ such that $\ell/N=o(1)$, where $C_\delta$ is a nonrandom constant.
\end{assumption}

Examples for the function $v(\cdot)$ as well as examples for the generation of the sequence $(a_j)_{j\in\{1,\ldots,Q\}}$ are given in \citet{hounyo2023wild}. Some of these examples will be utilized in the simulation study in Section \ref{simulation}.

In the next section, we present a wild bootstrap hypothesis test for $\hp_g$ for a fixed grouping index $g$. 
Subsequently, we will study all null hypotheses $\hp_1,\ldots,\hp_{d-1}$ simultaneously to find the selected grouping index $\hat{g}$ defined in \eqref{def:gstar}. 

\subsection{Fixed grouping}
Fix a grouping $g\in\{1,\ldots,d-1\}$. To test the null hypothesis $\hp_g$ given in \eqref{eq:nullhp0} against its two-sided alternative at significance level $\alpha\in (0,1)$, we define the (observed) test statistic as
\begin{equation*}
\tstat_g^{obs}=\sum_{t=1}^N \summ_{gt}. %\label{eq:sg}.
\end{equation*}
Under the assumption that the null hypothesis is true for the bootstrap samples, the bootstrap test statistics $\left(T_g^b: 1 \leq b \leq B\right)$ can be computed analogously, using (WBDD) pseudo-observations $\mathbf{U}_g^b$, $b=1,\ldots, B$ instead of the observations $\mathbf{U}_g$. To ensure that the bootstrap samples are obtained under the null (where the expected value of the sum of the observed values is equal to zero and thus the mean is zero), we center the observations around zero before generating bootstrap pseudo-observations.

Finally, the bootstrap $p$-values are given by
\begin{align*}
p_g= \frac{1}{B+1} \left(\sum_{b=1}^B \mathbf{1} \{|T_g^b| \geq |T_g^{obs}|\} +1\right)\qquad (g=1,\ldots,d-1), %\label{def:pvals}
\end{align*} in accordance with Eq.~\eqref{eq:p_boot}.

\subsection{All groupings}
To compute the selected grouping index $\hat{g}$ defined in Eq.~\eqref{def:gstar}, we need to test all hypotheses $\hp_1,\ldots, \hp_{d-1}$ simultaneously. In this, we aim to control the probability of rejecting at least one true null hypothesis, which is called the FWER. 
Different procedures can be used to control the FWER at level $\alpha$. If we assume that the p-values given in Eq.~\eqref{eq:p_boot} have positive dependency, we may use the \Sidak correction. In this case, we adjust the observed p-values according to the \Sidak correction, such that $p_{adj}=1-(1-p_{boot})^{d-1}$. The hypothesis $\mathcal{H}_g$ is rejected if and only if the corresponding adjusted p-value is at most $\alpha$. 
If we do not make any assumption about the dependency structure of the p-values, we may use the Bonferroni-Holm procedure. This procedure requires ordering the p-values in an ascending manner, such that $p_{(1)}\leq p_{(2)}\leq\ldots,\leq p_{(d-1)}$. The hypotheses corresponding to $p_{(1)},\ldots,p_{(i^\star)}$ are rejected, where $i^\star=\max\{i\in\{1,\ldots,d-1\}:p_{(j)}\leq \alpha/(d-j)\text{ for all }j=1,\ldots,i\}$. If $i^\star$ does not exist, no rejection is made. For more information, see \citet{dickhaus_2014}.
	
Controlling the FWER ensures that the probability of having one or more false rejections (where we falsely state that a certain grouping does not lead to separability) is at most $\alpha$.
	In the following simulation study, we compare both multiple testing procedures and investigate the influence of the WBDD parameters on the performance of the proposed procedure.

\section{Simulation Study}
\label{simulation}
In this section, we aim to assess the behavior of the proposed method described in Section \ref{seq:wbdd_proc} by applying it to the following three distinct time series.
	\begin{ex}\label{First.ex}
		For simulation studies, we have generated three times series $(y\in \mathbb{R}^t: t=1, \ldots, N = 50)$ according to a "signal plus noise" model of the form $y(t) = f(t) + \varepsilon(t)$, $t=1, \ldots, N$. For the signal part $f$, the following three choices have been considered.
		\begin{enumerate}
			\item  $f_{1}(t)=\sin(2\pi t/3)$\label{ex1}
			\item $f_{2}(t)= \exp(0.2 t)$\label{ex2}
			\item  $f_{3}(t)= 0.7\cdot\cos(\pi t/2)+0.5\cdot\cos(\pi t/3)$\label{ex3}.
		\end{enumerate}
		It is known that the true grouping indices are given by $g_1^\star=2$, $g_2^\star=1$, $g_3^\star=4$ for $f_1, f_2$ and $f_3$ respectively \citep{golyandina2018singular}.
		In all three cases, the noise term $\varepsilon(t), t=1, \ldots, N=50$ is independent and identically distributed with $\varepsilon(1)\sim N(0, Var(f)/SNR)$, with signal to noise ratio $SNR\in \{2, 5 \}$.
	\end{ex}
	The embedding step as well as the SVD step of the SSA pipeline have been carried out using the R package "Rssa" \citep{golyandina2018singular}. We have set the window length to $L=N/2$.
	To compute the grouping parameter as described above, we have employed the wild bootstrap method with $B = 1000$ bootstrap replications. % to thoroughly evaluate its robustness and effectiveness. 
	Implementation of the WBDD requires fixing the block size $\ell$, the weight function $v(t)$ as in Eq.~\eqref{eq:w} and the sequence of random variables $(a_j)_{j\in\{1,\ldots,Q\}}$. In accordance with \citet{hounyo2023wild}, we have set $\ell=N^{1/5}$. We have considered the two weight functions \begin{equation*}
			v^{[1]}(t)= \begin{cases}
			t &\text{if } t\in(0, 0.5]\\
			1-t &\text{if } t\in(0.5,1)\\
			0 & \text{else}, 
		\end{cases}
	\end{equation*}
	and
	\begin{equation*}
	v^{[2]}(t)=
	\begin{cases}
		t/0.43 &\text{if }t\in[0,0.43]\\
		1 &\text{if }t\in[0.43,1-0.43]\\
		(1-t/0.43) &\text{if }t\in[1-0.43,1]\\
		0&\text{if }t\notin[0,1],
	\end{cases}
	\end{equation*}
	where $v^{[2]}(t)$ corresponds to the example given in \citet{hounyo2023wild} with $c=0.43$.
	To investigate the influence of the external random sequence used in the WBDD on the estimated grouping index $\hat{g}$, we have utilized four different random sequences, namely those introduced in Examples 2.3-2.5 and 2.7 in \citet{hounyo2023wild}. We denote the  sequences by $a^{[i]}$, $i\in\{3,4,5,7\}$, where $i$ indicates the corresponding example in \citet{hounyo2023wild}.
	The grouping index $\hat{g}$ has then been obtained as described in Eq.~\eqref{def:gstar}. To account for multiplicity, we have used the \Sidak and Bonferroni-Holm correction with a significance level of $\alpha=0.1$. 
    
    We have compared the proposed procedure with the method for automatic grouping described by \citet{Bilancia2010} as implemented by \citet{golyandina2018singular}, Algorithm 2.15. This approach uses hierarchical clustering applied to the $w$-correlation matrix. The procedure groups the indices corresponding to the elementary matrices. We have set the grouping parameter returned by the hierarchical clustering approach to be the largest index in the first group.
    Simulation results are based on 500 Monte-Carlo iterations.\\
	
	In the simulation study, the choice of the weight function did not influence the results of the proposed procedure but the choice of the random sequence did. Furthermore, no differences in the results between utilizing the \Sidak or the Bonferroni-Holm correction have occurred.
	
	For all considered scenarios, both considered weight functions and all random sequences, the proportion of iterations for which $\hat{g}> g^\star$ has been below $\alpha=0.1$. Indeed, only for signal $f_2$ did the proportion differ from zero. In this case, the observed FWER has been lower for large $SNR$ than for low $SNR$. When comparing the observed FWER of the procedure utilizing the four different random sequences, the observed FWER has been the largest for $a^{[4]}$. Results for the grouping index returned by the hierarchical clustering approach are similar, only for $f_2$ and $SNR=2$ did the proportion of $\hat{g}> g^\star$ differ from zero. Still, the proportion has been below $\alpha=0.1$ in this scenario.

	Next, we comment on the accuracy of the proposed procedure. We say that the procedure is accurate if $\hat{g}$ is close to the true grouping index $g^\star$. 
    Table \ref{table:sim_mean_sd} displays the observed average and standard deviation of the values of the grouping index $\hat{g}$ based on the different random sequences $a^{[i]}$ as well as $\hat{g}$ based on the hierarchical clustering approach for the considered scenarios. We first focus on the results of the proposed procedure. Generally, accuracy has increased as the $SNR$ increases. Furthermore, utilizing the random sequence $a^{[4]}$ in the WBDD has lead to the most accurate average grouping indices for signals $f_1$ and $f_3$. Regarding the standard deviation, utilizing $a^{[4]}$ has lead to the largest standard deviation for signal $f_2$, for the other examples, $a^{[7]}$ and $a^{[5]}$ have lead to the largest standard deviations.
	Generally, the proposed procedure has been most accurate for signal $f_1$ and it has been least accurate for signal $f_3$. For signal $f_3$, the leading two signal components correspond to the trend, while the third and fourth signal components correspond to seasonality. Thus, the method appears to be somewhat insensitive to seasonal components of the signal.
    The automatic grouping approach based on hierarchical clustering has been either as accurate as or more accurate than the proposed procedure in all considered scenarios. The proposed procedure tends to be slightly more conservative (i.\ e., tending to smaller values of the selected grouping index), as it provides the theoretical guarantee that the probability of including noise in the reconstructed signal remains below $\alpha$.

\section{Real-world Data Analysis}
\label{real_world_examples}

This section demonstrates the application of the proposed procedure to disease surveillance data. A key task in this context is the extraction of underlying trends, as they provide a reliable picture of disease dynamics over time. However, raw data often contain substantial noise arising from reporting practices, calendar effects, and other short-term variations. Such noise can mask genuine patterns, lead to misleading interpretations, and affect forecasting. In contrast, isolating true signals provides a clear representation of disease progression and improves the capacity of the surveillance system to support public health decisions.

We have considered the daily series of Coronavirus Disease 2019 (COVID-19) hospitalizations recorded in the Emilia-Romagna region of Italy during 2022. The data were published by the Italian Civil Protection Department and are available at \url{https://github.com/pcm-dpc/COVID-19}. The time series consists of $N=365$ observations, which exhibit strong day-of-week effects, largely due to reporting delays over weekends that are subsequently corrected on Mondays and Tuesdays. We have applied our proposed procedure to extract the underlying trend of hospitalizations, in order to describe the evolution of the epidemic.

For the procedure to be valid, the window length $L$ must be fixed in advance. This parameter specifies the number of consecutive observations that are grouped in the embedding step of SSA and should approximate the local state of the underlying process \citep{hassani2007}. Its value should be proportional to eventual periodicities of the data and, for time series with complex structure, not too large in order to avoid an undesirable decomposition of the components of interest \citep{Golyandina2001}. Moreover, larger values of $L$ increase the number of tested hypotheses in our framework, producing generally higher adjusted $p$-values and thus a more conservative procedure, with the risk that the selected grouping $\hat{g}$ underestimates the optimal value $g^*$. Given the strong weekly structure of the series, we have set $L=7$, consistently with previous applications of SSA to COVID-19 data \citep{alharbi2021}. Furthermore, we have used the random sequence $a^{[4]}$, as given in Example 2.4 in \citet{hounyo2023wild}, and the Bonferroni-Holm correction with $\alpha=0.1$. In addition, we have computed the grouping parameter using the hierarchical cluster approach by \citet{Bilancia2010}.

\begin{figure}[h]
\begin{center}
	\includegraphics[height=3.6cm]{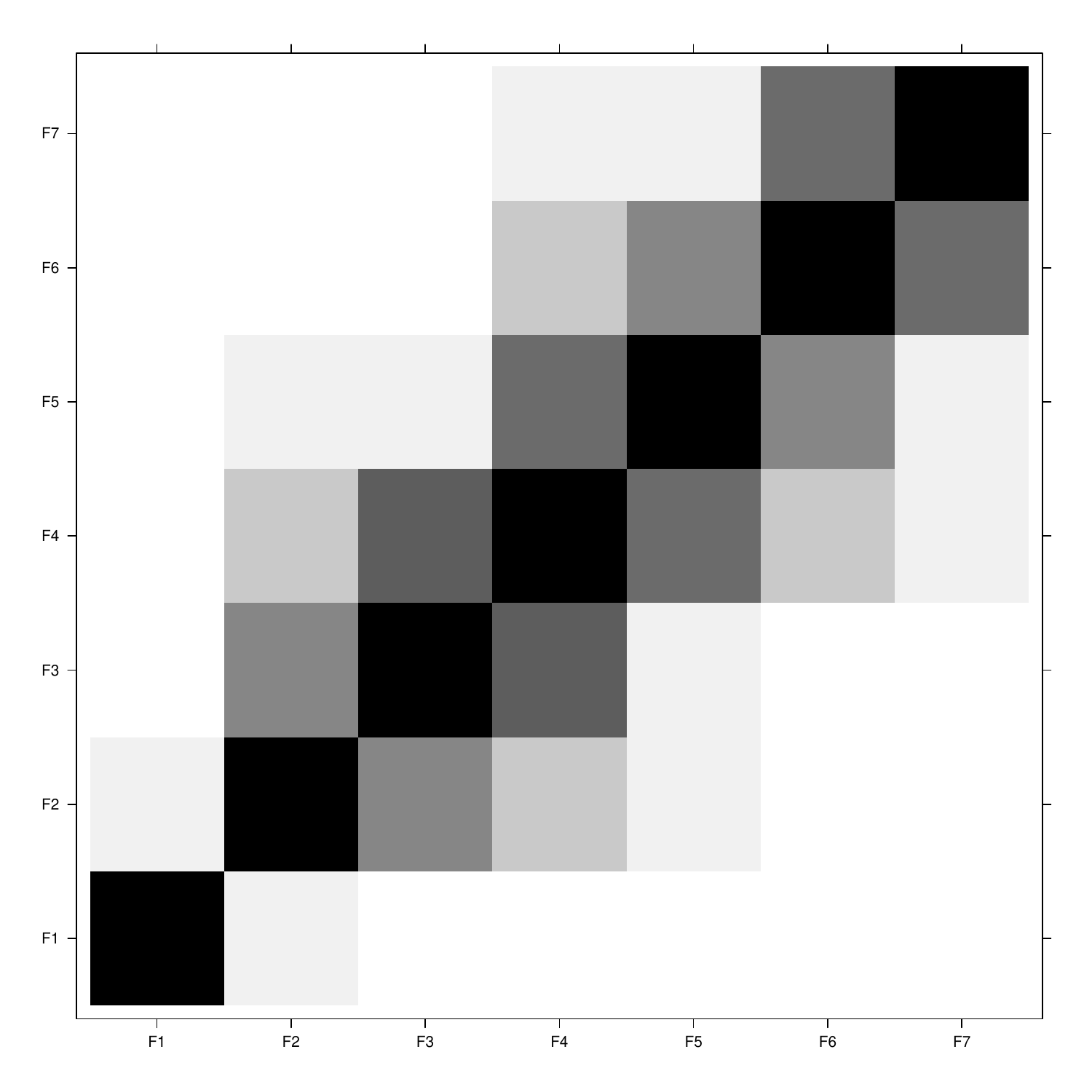}
    \hfill
    \includegraphics[height=3.6cm]{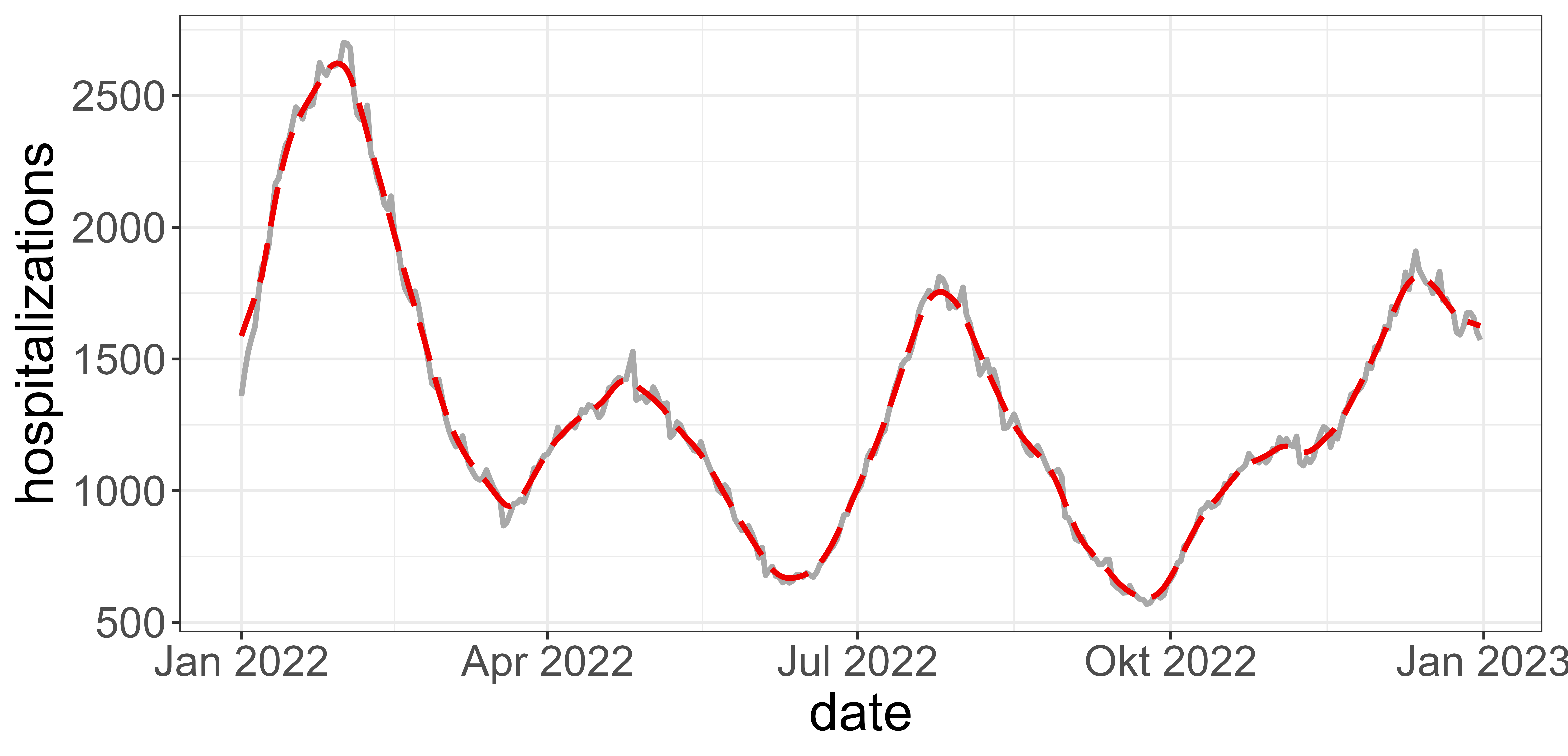}
	\caption{Illustration of SSA results: Analysis of COVID-19 hospitalizations in Emilia Romagna during 2022. The left plot displays the matrix illustration of the  ${w}$-correlations. The right plot shows the original time series (solid gray line) and the reconstructed signal based on the grouping according to the proposed procedure (dashed red line).}
	\label{f:case_study}
\end{center}
\end{figure}

Both procedures consistently identify the same grouping index $\hat{g}=1$. The $w$-correlation matrix (Figure \ref{f:case_study}, left) further supports this choice, as it shows separability among the components. Finally, the reconstructed signal (Figure \ref{f:case_study}, right) demonstrates a clear and accurate representation of the underlying structure, confirming the quality and reliability of the decomposition.

\section{Conclusion and Future Work}
\label{conclusion}
In this work, we have derived a procedure to identify the grouping index for separation of a time series into signal and noise in the context of SSA. In contrast to existing methods, the proposed procedure offers statistical guarantees in terms of a confidence statement regarding that no noise is included in the reconstructed signal.
We assume separability if the $w$-correlation is close to zero and utilize the wild bootstrap procedure proposed by \citet{hounyo2023wild} to approximate the null distribution of the empirical $w$-correlation.
The grouping index is determined by testing the null hypothesis of separability for all possible grouping indices and defining the selected grouping index $\hat{g}$ as the minimal index for which the null hypotheses cannot be rejected for all $g\in\{\hat{g},\ldots,d-1\}$. To account for the multiplicity of this approach, the Bonferroni-Holm correction can be used to control the FWER. 

We have investigated the performance of the proposed method in a simulation study. The results indicate that the proposed procedure returns accurate grouping indices if the signal is not complex, i.e., the true $g$ is low. Specifically, the method appears to be insensitive to seasonal components. The accuracy of the proposed procedure might increase for complex signals if a different window length is chosen. As pointed out by \citet{golyandina2018singular}, shorter window lengths might capture complex time series structures better.

Furthermore, the proposed procedure has been applied to disease surveillance data, i.e., the daily count of COVID-19 hospitalizations. Results were consistent with established methodologies, including the grouping method of investigating the plot of the $w$-correlation. Since the practitioner does not need to choose between multiple possible groupings, our proposal might increase the reproducibility of SSA studies. In addition, it provides statistical guarantees that trends or periodicity observed in the reconstructed signal likely do not include any noise.

This study opens avenues for future investigations. Further exploration could delve into refining the parameterization of the SSA method, investigating the impact of different parameter choices beyond $g$, and expanding the evaluation across diverse datasets. Moreover, extending this method's application to practical scenarios and different types of time series data would be valuable. Additionally, exploring variations or enhancements of the SSA method could lead to improved accuracy and broader applicability in various domains.
%In essence, the findings of this study underscore the potential of $\hat{g}$ and lay the groundwork for future research directions aimed at enhancing the performance and applicability of the SSA method in time series analysis.

\section{Acknowledgement}
Friederike Preusse gratefully acknowledges funding by the Deutsche Forschungsgemeinschaft (DFG, German Research Foundation)- project number 281474342. Anna Vesely acknowledges partial financial support by the Deutsche Forschungsgemeinschaft (DFG) via Grant No.~DI 1723/5-3, and from the Italian Complementary National Plan PNC-I.1 “Research initiatives for innovative technologies and pathways in the health and welfare sector” D.D.~931 of 06/06/2022, “DARE - DigitAl lifelong pRevEntion" initiative, code PNC0000002, CUP: B53C22006450001. We are grateful to Daniel Ochieng for his contributions to a previous version of this manuscript.

\section{Data availability}
%The data used in this article are available at the OpenNeuro dataset ds003871 at \url{https://openneuro.org/datasets/ds003871/versions/1.0.2}.
The data used in this article, provided and published by the Italian Civil Protection Department, are available at \url{https://github.com/pcm-dpc/COVID-19}.
The code for the simulation and its analysis is available upon request from the authors.

\section{Author Contribution}
MM and AV developed the methodology. TD recommended the usage of the wild bootstrap and wrote the introduction. AV and FP implemented the methodology and authored the corresponding sections. MM and FP performed the simulations. Application to real-world data was carried out by AV. All authors proofread and approved the final manuscript.
\newpage
\clearpage

\bibliography{sn-bibliography}
\newpage
\section{Tables}
 \begin{table}[h]
		\begin{tblr}{crrrrrr}
			&
            \SetCell[c=2]{c} Signal $f_{1}$
            && \SetCell[c=2]{c} Signal $f_{2}$ 
            && \SetCell[c=2]{c} Signal $f_{3}$ \\
			%\multicolumn{2}{c}{Signal $f_{2}$} &
			%\multicolumn{2}{c}{Signal $f_{3}$} \\
			$a^{[i]}$&$SNR=2$&$SNR=5$&$SNR=2$&$SNR=5$&$SNR=2$&$SNR=5$\\
			\hline
			$i=3$& 1.942 & 2 & 1.180  & 1.036 & 1.690 & 1.970 \\
			&  ($\pm$ 0.2340) & ($\pm$ 0)& ($\pm$ 1.0798) &  ($\pm$ 0.5722)&  ($\pm$ 0.5162)& ($\pm$0.3865)\\
			$i=4$ & 1.962& 2 & 1.230& 1.086  & 1.750  & 1.984 \\
			 & ($\pm$ 0.1914)& ($\pm$ 0)& ($\pm$ 1.2426)& ($\pm$ 0.8878) & ($\pm$ 0.5178) & ($\pm$ 0.3795)\\
			$i=5$ & 1.914 & 2 & 1.174  & 1.038 & 1.652& 1.908  \\
		 & ($\pm$ 0.2806)& ($\pm$ 0)& ($\pm$ 1.0854) & ($\pm$ 0.5264) &  ($\pm$ 0.5248) & ($\pm$ 0.4288) \\
			$i=7$ &  1.898  & 2 & 1.156 & 1.040 & 1.670 & 1.946\\
			 & ($\pm$ 0.3029) & ($\pm$ 0)& ($\pm$ 1.0649)& ($\pm$ 0.6318)&($\pm$ 0.5307)&($\pm$ 0.4092)\\
             \hline[dashed]
             HC& 2&2&1.008&1&2.212&2.124\\
             &($\pm$ 0)&($\pm$ 0)& ($\pm$ 0.0892) &($\pm$ 0) & ($\pm$ 0.6163) &($\pm$ 0.4828)\\
		\end{tblr}
		\caption{The average values of the selected grouping index $\hat{g}$ and the corresponding empirical standard deviations (in brackets) for different signals and varying $SNR$. The signals are defined in Example \ref{First.ex}.
        The proposed procedure has been based on the wild bootstrap for dependent data by \citet{hounyo2023wild} using four different random sequences $a^{[i]}$, $i\in\{3,4,5,7\}$, as given in Examples 2.3-2.7 in \citet{hounyo2023wild}. Additionally, the average grouping parameters based on the hierarchical clustering approach (HC) by \citet{Bilancia2010} and the corresponding standard deviations are reported.
        The true grouping index $g^\star$ corresponds to $g_1^\star=2$ for signal $f_1$, $g_2^\star=1$ for signal $f_2$ and $g_3^\star=4$ for signal $f_3$.
		 Results are based on 500 iterations.}
		\label{table:sim_mean_sd}
	\end{table}
\newpage
\section{Figures}

\end{document}